\documentclass[aps,twocolumn,amsfonts,showpacs,superscriptaddress]{revtex4} 
\usepackage[dvips]{graphicx}
\usepackage{bm}
\usepackage{epsfig,amsopn}
\usepackage{graphicx}
\usepackage{amsmath,amssymb}
\usepackage{natbib}
\newcommand{\1}{\mbox{$| 1\rangle$}}
\newcommand{\2}{\mbox{$| 2\rangle$}}
\newcommand{\3}{\mbox{$\langle 1 |$}}
\newcommand{\4}{\mbox{$\langle 2 |$}}
\newcommand{\sv}{\mbox{$| \psi(t)\rangle$}}
\newcommand{\svv}{\mbox{$| \varphi(t)\rangle$}}

\begin{document}

\preprint{APS/123-QED}

\title{Coherent destruction of tunneling, dynamic localization and the
Landau-Zener formula}

\author{Yosuke Kayanuma}
\affiliation{Department of Mathematical Sciences, Graduate School of Engineering \\
Osaka Prefecture University, Sakai 599-8531, Japan}
\author{Keiji Saito}
\affiliation{Department of Physics, Graduate School of Science \\ 
University of Tokyo, Bunkyo-ku, Tokyo 113-0033, Japan}

\date{\today}

\begin{abstract}
We clarify the internal relationship between the coherent destruction of 
tunneling (CDT) for a two-state model and the dynamic localization (DL) 
for a one-dimensional tight-binding model, under the periodical driving field. 
The time-evolution of the tight-binding model 
is reproduced from that of the two-state model by 
a mapping of equation of motion onto a set of ${\rm SU}(2)$ 
operators. It is shown that DL is effectively an infinitely large dimensional 
representation of the CDT in the ${\rm SU}(2)$ operators.
We also show that both of the CDT and the DL can be interpreted as a 
result of destructive interference in repeated Landau-Zener level-crossings.
\end{abstract}


\maketitle
The coherent control of quantum dynamics of electrons by a periodically 
oscillating external field has been one of the subjects 
of considerable interest both in nanoscale solid state physics\cite{Kohler}, 
and in molecular physics under laser fields\cite{Yamanouchi}. 
The interest is now extended to the trapped atoms in Bose-Einstein 
condensates\cite{Eckhardt}, 
the localized spins in molecular magnets\cite{Miyashita}, the Cooper pairs in 
Josephson qubits\cite{Sillampaa}, to name only a few. 
It should be noted that, even when the static properties of a quantum system is 
well known, its response to an explicitly time-dependent driving 
field may be nontrivial and, in some cases, poses a quite interesting 
problem. 
The phenomena known as the coherent destruction of tunneling (CDT)
\cite{Grossmann} 
and dynamic localization (DL)
\cite{Dunlap} are such typical nontrivial phenomena. 
Note that the CDT was originally found by Grossmann et al.\cite{Grossmann} 
for a model of double-well potential, but it has been made clear 
that the essential mechanism of the phenomenon can be well understood by
a two-state model which represents the quantum dynamics between the lowest
two states localized to each well \cite{gross_euro,gomez}

In both CDT and DL, the initial localized quantum state never diffuses 
under a periodic external field.
In this aspect, these phenomena are similar 
However, there are also some dissimilarities.
The DL is an exact result obtained in an infinite driven system and is
valid irrespective of the magnitudes of the transfer matrix element. 
On the other hand, the CDT is derived approximately in an extreme case of 
a small value of the transfer matrix element. In the CDT, the initial 
distribution is frozen, but in the DL, the distribution oscillates 
around the initial value. Thus 
the relation between the CDT and the DL has been 
controversial\cite{Grifoni,Raghavan}.

In this Rapid Communication, we study the relationship 
between these remarkable phenomena in a unified way.
One may study this problem by assuming 
a tight-binding model with a finite length, and by observing the change of the 
behavior of the electron according to the change of the 
chain length\cite{Raghavan}. 
However, the equation of motion for a finite linear chain model does not allow 
for the analytical solution, so that 
the analysis inevitably becomes a numerical one. 
We present here a new approach to this problem, which shed 
light upon the internal relationship between the CDT and the DL. 
It will be shown that the DL is an infinitely large dimensional 
representation of a generalized version of the CDT, and in fact they are 
closely related each other. 
 
\par
Let the 
two-state system \1 and \2 be under an external field and driven by the 
Hamiltonian,
\begin{equation}
H_1(t)=\frac{E(t)}{2}\left(\1\3-\2\4\right)+\gamma\left(\1\4+\2\3\right),
\label{first_hamiltonian}
\end{equation}
where $\gamma$ is a constant tunneling matrix element. The Schr\"odinger 
equation $(\hbar=1)$,
$
i{d/dt}\sv =H_1(t)\sv
$, 
is cast into the form
\begin{eqnarray}
i{d\over dt}a_1(t)&=&\frac{E(t)}{2}a_1(t)+\gamma a_2(t),\nonumber \\
i{d\over dt}a_2(t)&=&-\frac{E(t)}{2}a_2(t)+\gamma a_1(t), \label{a_representation}
\end{eqnarray}
in the representation $\sv=a_1(t)\1 + a_2(t)\2$.
Although this is the simplest equation of quantum dynamics, it cannot be solved 
analytically for general functional forms of $E(t)$. 
Grossmann and H\"{a}nggi \cite{gross_euro} and Llorente and Plata \cite{gomez}
pointed out that for a sinusoidal 
time-dependence of the driving field $E(t)=E_0\cos(\omega t)$, Eq.(\ref{a_representation}) is solved approximately in the limit of rapid modulation 
$\omega \gg\gamma$. By substituting $a_1(t)=\exp[-i(E_0/2\omega)\sin(\omega t)]c_1(t)$, $a_2(t)=\exp[i(E_0/2\omega)\sin(\omega t)]c_2(t)$, 
Eq.(\ref{a_representation}) is rewritten as
\begin{eqnarray}
i{d\over dt}c_1(t)&=&\gamma\exp[i(E_0/\omega)\sin(\omega t)]c_2(t),\nonumber \\
i{d\over dt}c_2(t)&=&\gamma\exp[-i(E_0/\omega)\sin(\omega t)]c_1(t).
\label{c_representation}
\end{eqnarray}
In the limit $\omega \gg\gamma$, the above equation 
is integrated approximately for a short period $2\pi/\omega$ 
by assuming that $c_1(t)$ and $c_2(t)$ are constant, since the rapidly 
oscillating terms are separated out as the phase factors. 
This is the inverse adiabatic approximation. We obtain 
\begin{eqnarray}
i{d\over d\tau }c_1(\tau )&=&\gamma J_0(E_0/\omega)c_2(\tau ),\nonumber \\
i{d\over d\tau }c_2(\tau )&=&\gamma J_0(E_0/\omega)c_1(\tau ), \label{rapid_case_equations}
\end{eqnarray}
where $\tau$ is 
a coarse-grained time by the unit of $2\pi/\omega$, and 
$J_0(E_0/\omega)$ is the zeroth order Bessel function:
$$
J_0(E_0/\omega)\equiv {\omega\over 2\pi}\int_t^{t+2\pi/\omega}\exp[i(E_0/\omega)\sin(\omega u )]d u
$$
The above equation tells us that the tunneling parameter is reduced 
effectively by the factor $J_0(E_0/\omega)$, and even vanishes 
in the case that $E_0/\omega$ coincides with a zero of the Bessel function. 
This is the CDT\cite{Grossmann}.
\par
An infinite dimensional analogue of the model (1) is given by the Hamiltonian
\begin{eqnarray}
H_2(t)&=&E(t)\sum_{n=-\infty}^\infty n|n\rangle\langle n|\nonumber \\
&+&\Delta \sum_{n=-\infty}^\infty \left(|n\rangle\langle n+1|+|n+1\rangle\langle n|\right). \label{second_hamiltonian}
\end{eqnarray}
This is a model Hamiltonian for an electron in an infinite one-dimensional chain under a time-dependent electric field, where $|n \rangle$ represents the Wannier state at 
site $n$. Paradoxically, 
the Schr\"odinger equation
$i{d/dt}\svv=H_2(t)\svv$ is solved analytically {\it for arbitrary 
functional forms of $E(t)$}. 
We show explicit time-evolution operator with a Lie algebra.
We define $T_0=\sum_{n=-\infty}^\infty n|n\rangle\langle n|$, $T_+=\sum_{n=-\infty}^\infty |n+1\rangle\langle n|$, $T_-=\sum_{n=-\infty}^\infty |n\rangle\langle n+1|$ to get $H_2(t)=E(t)T_0+\Delta\left(T_++T_-\right)$.
These operators satisfies the relations:
\begin{equation}
[T_0,T_\pm ]=\pm T_\pm,\quad
[T_+, T_- ]=0.
\end{equation}
The solution of the Schr\"odinger equation is written as 
$
\svv=U(t)|\varphi(0)\rangle,
$
where
$
U(t)=\exp_+(-i\int_0^t H_2(s) ds ).
$
By Feynman's disentangling theorem, $U(t)$ is written in the form,
\begin{equation}
U(t)=e^{-iA(t)T_0}\exp_+\left[-i\Delta\int_0^t\left(\tilde T_+(u)+\tilde 
T_-(u)\right)du \right], \nonumber
\end{equation}
where $A(t)\!\equiv \!\int_0^tE(u)du$, and $
\tilde T_\pm(u) \!\equiv \! e^{iA(u)T_0}T_\pm e^{-iA(u)T_0}
\!= \! e^{\pm iA(u)}T_\pm.
$
Since $T_+$ and $T_-$ are commutable, $U(t)$ is rewritten as
\begin{equation}
U(t)=\exp[-iA(t)T_0]\exp[-i B(t)] \label{U_representation}
\end{equation}
in which $B(t)=\Delta\left\{R(t)T_++R(t)^*T_-\right\}$ with
$R(t)=\int_0^t\exp\left[iA(u)\right]du$. Since $B (t)$ has the 
translational symmetry, its eigenstates are given by 
the plane waves $|k\rangle = \sum_n e^{ikn}|n\rangle$ with the time-dependent 
eigenvalue 
$\epsilon_k(t)=\Delta\left\{R(t)e^{-ik}+R^*(t)e^{ik}\right\}$. Then the matrix element for the transition $|n\rangle \rightarrow |m\rangle$ 
is calculated by using the closure relation as
\begin{eqnarray}
\langle m|U(t)|n\rangle &=&
\exp\left[-iA(t)m+i\left(\chi +\frac{\pi}{2}\right)(m-n)\right]
\nonumber \\
&\times& J_{m-n}\left(2\Delta|R(t)|\right), \label{closure_form}
\end{eqnarray}
where $\chi=\arg R(t)$ and $J_n(x)$ is the $n$th order Bessel function. For 
a specific choice $E(t)=E_0\cos(\omega t)$, and at each period 
of the oscillation $\tau=2\pi l/\omega\quad (l=0,1,2,\cdots)$, 
we find $A(\tau ) =0$ and $R(\tau ) = \tau J_0 (E_0 / \omega )$, and 
the transition probability is given by
\begin{equation}
|\langle m|U(\tau )|n\rangle |^2
=
J_{m-n}^2 \left(2\tau \Delta 
|J_0(E_0/\omega)|
\right) .
\label{DL_probability}
\end{equation}
This should be compared with the value $J_{m-n}^2
\left(2\tau \Delta \right)$ 
which corresponds to the case without external field. 
Eq.(\ref{DL_probability}) 
indicates that the oscillating external field generally reduces 
the effective transfer by the factor $J_0(E_0/\omega)$. Especially, 
if $E_0/\omega$ coincides with a zero of $J_0(x)$, the probability to find 
the electron at site $m(\neq n)$ oscillates temporally and becomes zero,  
while that to find it at the original site $n$ becomes unity at each period 
$2\pi/\omega$. This is the dynamic localization (DL)\cite{Dunlap}. 

It is clear that the integrability of the Schr\"odinger equation 
for (\ref{second_hamiltonian}) rests upon the commutativity 
of $T_+$ and $T_-$. On the other hand,
for the two-state model (\ref{first_hamiltonian}), we can  
define the analogous operators, 
$S_0=\frac{1}{2}\left(\1\3-\2\4\right)$, $S_+=\1\4$, and $S_-=\2\3$,
to get $H_1 (t) =E(t) S_0 + \gamma (S_++S_-)$.
These operators, however, satisfy a true ${\rm SU}(2)$ Lie algebra:
\begin{eqnarray}
[S_0, S_\pm]=\pm S_\pm,\quad [S_+, S_-]=2S_0. 
\end{eqnarray} 
These are uncommutable relations
and the time-evolution operator cannot be decomposed in general.

\par
Let us discuss the relations between the dynamics of CDT and DL. 
We consider the following bosonic representation 
for (\ref{first_hamiltonian}) with Schwinger bosons:
\begin{equation}
H_3(t)=\frac{E(t)}{2}\left(b_1^\dagger b_1 -b_2^\dagger b_2\right) 
+ \gamma \left(b_1^\dagger b_2+b_2^\dagger b_1\right),
\end{equation}
where $b_i$ satisfies the commutation relation of independent bosons,
$[b_i, b_j^\dagger]=\delta_{i,j}$. 
The Heisenberg equation for $b_1^\dagger$ and $b_2^\dagger$ is given by
\begin{eqnarray}
i{d\over dt}b_1^\dagger(t)&=&-\frac{E(t)}{2}b_1^\dagger(t)-\gamma b_2^\dagger(t),\nonumber \\
i{d\over dt}b_2^\dagger(t)&=&\frac{E(t)}{2}b_2^\dagger(t)-\gamma b_1^\dagger(t),
\label{b_representation}
\end{eqnarray}
which is equivalent to Eq.(\ref{a_representation}) by the replacement
$b_{i}^{\dagger} (t)$ by $a_{i} (t)$. 
The solution of Eq.(\ref{b_representation}) with the initial conditions, 
$b_1^\dagger(0)=b_1^\dagger$, and $b_2^\dagger(0)=b_2^\dagger$ is 
generally written as
\begin{eqnarray}
\left(\begin{array}{c} b_1^\dagger (t)\\ b_2^\dagger (t)\end{array}\right)
&=& U_b \left(\begin{array}{c} b_1^\dagger \\ b_2^\dagger \end{array}\right)\\
U_b&=&\left(\begin{array}{cc} \alpha & \beta\\
-\beta^* & \alpha^* \end{array}\right) \nonumber 
\end{eqnarray}
where $\alpha$ and $\beta$ are time-dependent complex numbers 
satisfying $|\alpha|^2+|\beta|^2=1$. The point is that, if the two-state dynamics described by Eq. (2) is solved 
somehow, it can be mapped onto the solution for Eq. (12) and we obtain 
a class of solutions for state vectors in higher dimensional representation spaces of SU(2). Recently, Pokrovsky and Sinitsyn\cite{Pokrovsky} 
utilized the same argument to derive a class of exact formulas describing the nonadiabatic transitions 
for a model of multiple level crossings.
\par

Let us define the basis states designated by the boson numbers,
\begin{equation}
|\Psi\rangle = |p,q\rangle =
\frac{1}{\sqrt{p!q!}}{b_1^\dagger}^p{b_2^\dagger}^q|{\rm vac}\rangle,
\end{equation}
where $|{\rm vac}\rangle$ is the vacuum state of the bosons. 
The total boson number is a constant of motion. We fix $p+q=2N$, and define the {\it site} index $n$ by  $n\equiv (p-q)/2$. The basis 
states are classified as $|n\rangle =|N+n,N-n\rangle, (n=-N, -N+1, \cdots, 0, 1, \cdots, N)$. The nonzero off-diagonal 
matrix elements are then given by
$
\langle n+1|H_3|n\rangle =\gamma \sqrt{(N+n+1)(N-n)}.
$
If we set $\gamma=\Delta/N$, we have a $2N+1$-dimensional linear chain model as a representation of the ${\rm SU}(2)$ Hamiltonian,
\begin{eqnarray}
H_3(t)&=&E(t)\sum_{n=-N}^N n|n\rangle\langle n|\nonumber \\
&+&\Delta \sum_{n=-N}^{N-1}f_n\left(|n+1\rangle\langle n|+|n\rangle\langle n+1|\right), \label{third_hamiltonian}
\end{eqnarray}
in which $f_n=\sqrt{\left(1+\frac{n+1}{N}\right)\left(1-\frac{n}{N}\right)}$. Also in the sector $p+q=2N-1$, an analogous expression is 
obtained. Specifically, for $p+q=1$, the two-state model $H_1(t)$ is recovered. An important observation here is that, in the limit 
$N\rightarrow \infty$ with fixed $n$, the tight-binding model with an infinite chain $H_2(t)$ is also 
recovered since $f_n\rightarrow 1$. 
Thus the CDT dynamics in $U_{b}$ can be connected to the DL dynamics
in the wave functions for the Hamiltonian (\ref{third_hamiltonian}).

We now study the time-evolution operator for the wave function, $V(t)$, 
which satisfies,
$
|\Psi(t)\rangle = V(t)|\Psi\rangle
$
Once explicit matrix elements in $U_b$ are obtained, 
one obtains a class of time-evolutions for the driven system 
(\ref{third_hamiltonian}). 
The wave function $|\Psi (t)\rangle$ is given with $U_b$ as
\begin{equation}
|\Psi(t)\rangle =\frac{1}{\sqrt{p!q!}}\left(\alpha^*b_1^\dagger - \beta b_2^\dagger\right)^p
\left(\beta^* b_1^\dagger +\alpha b_2^\dagger\right)^q |{\rm vac}\rangle. ~~~~~~
\end{equation}
By expanding the right hand side, and rearranging the terms proportional to ${b_1^\dagger}^{N+m} {b_2^\dagger}^{N-m}$, we find the 
transition amplitude for $|n\rangle \rightarrow |m\rangle$, 
\begin{eqnarray}
\langle m|V(t)|n\rangle &=&\sqrt{\frac{(N+m)!(N-m)!}{(N+n)!(N-n)!}}{\alpha^*}^{n+m}{\beta^*}^{m-n} \nonumber \\
&\times&\sum_{r=r_m}^{r_M}\left(\begin{array}{c}N+n\\r\end{array}\right)\left(\begin{array}{c}N-n\\N-n-r\end{array}\right) 
\nonumber \\
&\times&|\alpha|^{2(N-m-r)}(-|\beta|^2)^r,
\end{eqnarray}
where the summation over $r$ runs from $r_m=\max\{0, n-m\}$ to $r_M=\min\{N+n,N-m\}$. 
This is rewritten as, 
\begin{eqnarray}
&&\langle m|V(t)|n\rangle \nonumber \\
&&~~~=\sqrt{\frac{(N+m)!(N-m)!}{(N+n)!(N-n)!}}{\alpha^*}^{m+n}{\beta^*}^{m-n}
P_{N-m}^{m-n,m+n}(x), \nonumber \\
\label{vt_expression}
\end{eqnarray}
where $x=2|\alpha|^2-1$, and $P_{N-m}^{m-n,m+n}(x)$ is Jacobi's polynomial\cite{Landau-Lifshitz}
defined as,
\begin{equation}
P_n^{(a,b)}(x)=\frac{1}{2^n}\sum_{r=0}^n \left(\begin{array}{c}n+a\\n-r\end{array}\right)\left(\begin{array}{c}n+b\\r\end{array}\right)(x-1)^r (x+1)^{n-r}.
\nonumber
\end{equation}
This expression of $V(t)$ is valid in the region $m-n\geq 0,\  m+n\geq 0$. In other regions, $\langle m|V(t)|n\rangle $ is given by the replacement; $m\rightarrow -n,\  n\rightarrow -m, \ \alpha^*\rightarrow \alpha$ for $m-n\geq 0,\  m+n\leq 0$,\  $m\rightarrow n, \ n\rightarrow m, \ \beta^*\rightarrow -\beta$ for $m-n\leq 0,\  m+n\geq 0$, and 
$m\rightarrow -m, \ n\rightarrow -n\  \alpha^*\rightarrow \alpha, \ \beta^*\rightarrow \beta$ for $m-n\leq 0, \ m+n\leq 0$. 
\par
Now set $E(t)$ to a sinusoidal modulation, $E(t)=E_0\cos(\omega t)$ with $\gamma=\Delta/N$. The condition for the rapid modulation limit $\omega \gg\gamma$
is satisfied for Eq. (\ref{b_representation}) in the limit $N\gg1$, and it is solved just the same way as the corresponding equation for the c-numbers (\ref{a_representation}). 
We obtain 
\begin{eqnarray}
\alpha(\tau )&= &\exp\left[i\frac{E_0}{2\omega}\sin\omega \tau\right]
\cos\left(\frac{\Delta}{N}J_0\left(E_0/\omega\right)\tau \right),\nonumber\\
\beta(\tau )&=&i\exp\left[i\frac{E_0}{2\omega}\sin\omega \tau \right]
\sin\left(\frac{\Delta}{N}J_0\left(E_0/\omega\right)\tau \right). 
\label{alpha_beta_expression}
\end{eqnarray}
Note that the time $\tau$ is coarse-grained by the unit $2\pi/\omega$. 
The following formula is easily proved by using Stirling's formula\cite{Gradshteyn},
\begin{equation}
\lim_{N\rightarrow\infty} N^{-a}P_N^{(a,b)}\left(1-\frac{z^2}{2N^2}\right)=\left(\frac{z}{2}\right)^{-a}J_a(|z|).
\end{equation}
Then, inserting Eq.(\ref{alpha_beta_expression}) 
into Eq.(\ref{vt_expression}), and noting that 
$x=\cos\left(2\Delta/N J_0(E_0/\omega)t\right)$, we get, in the limit $N\rightarrow\infty$,
\begin{eqnarray}
\langle m|V(\tau )|n\rangle &=&
\exp \left[ -i {E_0 m \over \omega } \sin \omega \tau + 
i{\pi \over 2}(m-n) \right] \nonumber \\
&& \times 
J_{m-n}\left(2\Delta \tau  \left|J_0\left(E_0/\omega\right)\right| \right).
\end{eqnarray}
This is exactly the same as the formula (\ref{closure_form}) 
including the phase factor. Especially, when $E_0/\omega$ coincides with a  
zero of $J_0( E_0/\omega )$, the CDT occurs in $U_b$, while the DL 
occurs in $V(t)$. Thus 
it is shown that the DL is an infinitely large dimensional 
representation of a generalized version of the CDT.
\par
One of the special cases of a class of the Hamiltonian 
(\ref{first_hamiltonian}) that allows 
the exact solution is the Landau-Zener model\cite{Landau,Zener} 
$E(t)=vt$. The solution is written in terms of Weber functions, and 
the transition probability from one branch to another according to the 
temporal evolution from $t=-\infty$ to $t=\infty$ is given by the 
celebrated Landau-Zener formula\cite{Landau,Zener}. 
One of the present authors\cite{Kayanuma} pointed out that CDT can be 
regarded as a result of destructive interference between the transition paths 
for repeated Landau-Zener level crossings. The above result suggests the 
possibility to extend this 
view to the DL. In the case $E(t)=vt\  (v>0)$, the transition matrix 
elements without adiabatic phases are given by
\begin{equation}
\alpha=\sqrt{P}, \quad \beta=-\sqrt{1-P}e^{i\phi}
\end{equation}
where $P=\exp\left[-2\pi\delta\right]$ is the Landau-Zener nonadiabatic transition probability with $\delta=\Delta^2/(N^2 v)$, and 
$\phi$ is the Stokes phase given by $\phi=\pi/4+\arg \Gamma(1-i\delta)+\delta(\ln \delta -1)$ in which $\Gamma(z)$ is the $\Gamma$ function. 
The transfer matrix for the two-state Landau-Zener model can be mapped onto the $2N+1$-site representation $S$ as before. 
Noting that, in the limit $N\gg 1$, $\alpha^2\simeq 1-2\pi\delta$ and $\beta\simeq -\sqrt{2\pi\delta}e^{i\pi/4}$, we find for the 
matrix element $\langle m|S|n\rangle $ at the crossing 
\begin{equation}
\langle m|S|n\rangle = \exp\left[-i\frac{\pi}{4}(m-n)\right] J_{m-n}\left(2\sqrt{2\pi\Delta/v}\right).
\end{equation}
This formula agrees with the exact formula obtained from 
Eq.(\ref{closure_form}) , as it should. 
\par
For a repeated crossings of the two-state model driven by $E(t)=E_0\cos(\omega t)$, and in the case that $E_0$ is much larger than $\gamma$ and $\omega$, 
we can approximately decompose the whole process into sudden transitions at level-crossings and the free propagation between them\cite{Kayanuma}. 
The velocity of energy change $v$ is given by the value estimated at the crossings, $v=E_0\omega$.
This is also mapped onto the $2N+1$-dimensional representation. 
Thus, for a double crossing within a period of the oscillation, say at $t_1=\pi/2\omega$ and $t_2=3\pi/2\omega$, we have the transition amplitude 
$\langle m|T|n\rangle$ from $|n\rangle$ to $|m\rangle$ in the limit $N\rightarrow \infty$ as a sum of all contribution from the intermediate states,
\begin{equation}
\langle m|T|n\rangle=\sum_{l=-\infty}^\infty \langle m|S|l\rangle e^{-i\Omega l}\langle l|S^T|n\rangle,
\end{equation}
where $S^T$ is the transpose of $S$, and $\Omega\equiv \int_{t_1}^{t_2}dt E_0\cos\left(\omega t\right)= 2E_0/\omega$. 
The summation is carried out exactly by using Graf's formula\cite{Watson},
\begin{eqnarray}
&&\sum_{m=-\infty}^\infty J_{\nu+m}(z)J_m(\zeta)e^{im\theta}\nonumber \\
&&~~=J_\nu\left(\sqrt{z^2+\zeta^2-2z\zeta\cos\theta}\right) 
\left\{\frac{z-\zeta e^{-i\theta}}{z-\zeta e^{i\theta}}\right\}^{\nu/2},
\nonumber
\end{eqnarray}
valid for real numbers $z$ and $\zeta$. 
The transition probability is thus obtained as
\begin{eqnarray}
|\langle m|T|n\rangle|^2
&=&J_{m-n}^2\left[2\Delta\sqrt{2\omega/\pi E_0}\sin\left(\frac{E_0}{\omega}+\frac{\pi}{4}\right)\frac{2\pi}{\omega}\right]. \nonumber \\
\end{eqnarray}
If one notices the asymptotic formula $J_0(x)\simeq \sqrt{2/\pi x}\sin\left(x+(\pi/4)\right)$ for $x\gg 1$, it can be seen that 
the formula (\ref{DL_probability}) agrees at $t=2\pi/\omega$ with the above one in the limit $E_0/\omega \gg 1$. The phase factor $\pi/4$ is nothing but the 
Stokes phase at the level-crossing in the diabatic limit. 
Thus it is revealed that, in the level of the two-state model, the CDT is a result of interference 
between the two intermediate transition paths, while in its infinitely large dimensional representation, 
the DL is a result of interference between infinite number of intermediate transition paths. 
\par 
This work was supported by the Grant-in-Aid from the Ministry of Education, Science, Sports, Culture and Technology of Japan (No. 18540323).


\begin{thebibliography}{notitle}
\bibitem{Kohler} See for example, S. Kohler, J. Lehmann, and P. H\"anggi, Phys. Rep. 406, 379 (2005), and references cited therein.
\bibitem{Yamanouchi} {\it Progress in Ultrafast Intense Laser Science} I, edited by K. Yamanouchi, S. L. Chin, P. Agostini and G. Ferrante, 
(Springer, Heidelberg, 2006), and A. D. Bandrauk and H. Kono in {\it Advances in Multiphoton Processes and Spectroscopy}, Vol. 15, edited by S. H. Lin, 
A. A. Villaeys, and Y. Fujimura, (World Scientific, Singapore, 2003), pp.147-214.
\bibitem{Eckhardt} A. Eckhardt, C. Weiss, and M. Holthaus, Phys. Rev. Lett. {\bf 95}, 260404 (2005).
\bibitem{Miyashita} S. Miyashita, K. Saito, and H. De Raedt,Phys. Rev. Lett. {\bf 80}, 1525 (1998).
\bibitem{Sillampaa} M. Sillanp\"a\"a, T. Lehtinen, A. Paila, Y. Makhlin, and P. Hakonen, Phys. Rev. Lett. {\bf 96}, 187002 (2006).
\bibitem{Grossmann} F. Grossmann, T. Dittrich, P. Jung, and P. H\"anggi, Phys. Rev. Lett. {\bf 65}, 2927 (1990).
\bibitem{Dunlap} D. H. Dunlap and V. M. Kenkre, Phys. Rev. B {\bf 34}, 3625 (1986).
\bibitem{gross_euro} F. Grossmann and 
P. H\"anggi, Europhys. Lett. {\bf 85}, 571 (1992).
\bibitem{gomez} J. M. Gomez Llorente and J. Plata, Phys. Rev. A {\bf 45},
R6958 (1992).
\bibitem{Grifoni} M. Grifoni and P. H\"anggi, Phys. Rep. {\bf 304}, 229 (1998).
\bibitem{Raghavan} S. Raghavan, V. M. Kenkre, D. H. Dunlap, A. R. Bishop, and M. I. Salkola, Phys. Rev. A {\bf 54}, R1781 (1996).
\bibitem{Pokrovsky} V. L. Pokrovsky and N. A. Sinitsyn, Phys. Rev. B {\bf 69}, 104414 (2004).
\bibitem{Landau-Lifshitz} L. D. Landau and E. M. Lifshitz, {\it Quantum Mechanics}, (Pergamon, Oxford, 1976), Chap. 8.
\bibitem{Gradshteyn} I. S. Gradshteyn and I. M. Ryzhik, {\it Tables of Integrals Series and Products}, (Academic Press, New York, 1965), Chap. 8.
\bibitem{Landau} L. Landau, Phys. Z. Sowjetjunion {\bf 2}, 46 (1932).
\bibitem{Zener} C. Zener, Proc. R. Soc. London, Ser. A {\bf 137}, 696 (1932).
\bibitem{Kayanuma} Y. Kayanuma, Phys. Rev. A {\bf 50}, 843 (1994).
\bibitem{Watson} G. N. Watson, {\it A Treatise on the Theory of Bessel Functions}, (Cambridge University Press, Cambridge, 1922), Chap. 11.



\end{thebibliography}
\end{document}